# A Survey of Active Attacks on Wireless Sensor Networks and their Countermeasures

Furrakh Shahzad[1], Maruf Pasha[2], Arslan Ahmad[2]

([1]Department of Computer Science, Pakistan Institute of Engineering and Technology, Multan 60000, Pakistan)
([2]Department of Information Technology, Bahauddin Zakariya University, Multan 60000, Pakistan)

farrukhshahzad@piet.edu.pk, maruf.pasha@bzu.edu.pk, arslan_ahmad91@yahoo.com

*Abstract*— **Lately, Wireless Sensor Networks (WSNs) have become an emerging technology and can be utilized in some crucial circumstances like battlegrounds, commercial applications, habitat observing, buildings, smart homes, traffic surveillance and other different places. One of the foremost difficulties that WSN faces nowadays is protection from serious attacks. While organizing the sensor nodes in an abandoned environment makes network systems helpless against an assortment of strong assaults, intrinsic memory and power restrictions of sensor nodes make the traditional security arrangements impractical. The sensing knowledge combined with the wireless communication and processing power makes it lucrative for being abused. The wireless sensor network technology also obtains a big variety of security intimidations. This paper describes four basic security threats and many active attacks on WSN with their possible countermeasures proposed by different research scholars.**

***Keywords**-Wireless Sensor Network (WSN) threats; Security Attacks; Security Goal; Wireless Network Challenges; Defensive mechanisms*

## I. INTRODUCTION

The significant improvements of hardware engineering methods and effectual software procedures create a network composed of several affordable sensors called a wireless sensor network [1-3]. WSN offers an auspicious network arrangement for different applications like home appliance management, environmental monitoring and medical care. This is widely used in homeland security and battlefield surveillance scenarios because WSNs are simple to install and effective for such circumstances [4]. However, in numerous tactical and hostile situations as well as significant vendible applications, safety appliances are predominantly required to secure the WSN from the malevolent threats [5]. Therefore, safety in WSN becomes a crucial and challenging task. Some applications of the wireless sensor network are also shown in fig.1.

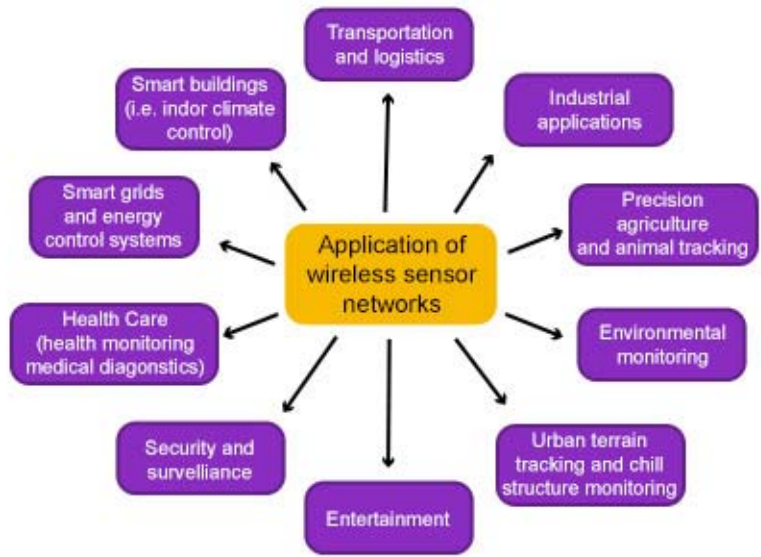

Fig. 1. Wireless Sensor Network (WSN) applications

A WSN usually contains a big amount of sensor devices where each sensor device is inadequate in its capabilities because of the price anxiety during the manufacturing process. For instance, MICA2 MPR400CB [6], a most prevalent platform of sensor nodes, has about 128 Kilo-Byte program storage and 8-bit ATmega128L Microchip [7]. In different circumstances, it is totally worthless how wisely a safety mechanism for WSN is designed, intruders can still become successful to find the vulnerability and perform attacks on the targeted node in WSNs. If intruders remain silent to eavesdrop on the network traffic, they can be remained hidden. If they try to disturb the network communications, irregularities and anomalies will be identified. Interrupt detection systems can notice these attacks by observing the behavioral change and anomalies in the network. A simple type of Wireless Sensor Network is depicted in fig. 2.

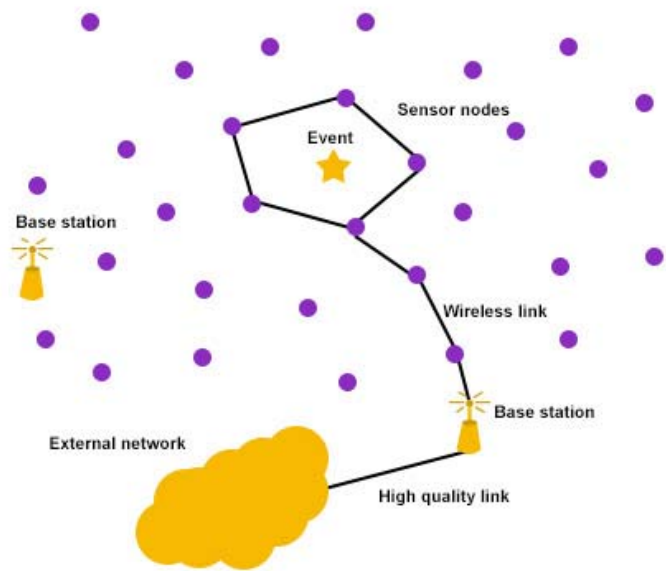

Fig. 2. Wireless Sensor Network (WSN)

*1.1 Some Sensor Networks Applications*

There are different kinds of sensors in wireless sensor network, for instance, thermal, seismic, visual, low sampling rate magnetic, acoustic, radar and infrared. Moreover, a large variety of conditions are monitored by sensor devices which include temperature, humidity, lightening condition, vehicular movement, Pressure, noise levels, the absence or presence of some objects kind, soil makeup, current characteristics like direction, speed, object size and mechanical levels of stress on objects that are attached [31]. Three Generations of Sensor network nodes and attributes of sensor networks are shown in table 1 & table 2 respectively.

Table 1
THREE GENERATIONS OF SENSOR NODES

|  | 1980's-1990's | 2000-2003 | 2010 |
|---|---|---|---|
| Manufacturer | Custom Contractors, e.g., for TRSS | Commercial: Crossbow Technology, Inc. Sensoria Corp., Ember Corp. | Dust, Inc. and others to be formed |
| Size | Large shoe box and up | Pack of cards to small shoe box | Dust particle |
| Weight | Kilograms | Grams | Negligible |
| Node architecture | Separate sensing, processing and communication | Integrated sensing, processing and communication | Integrated sensing, processing and communication |
| Topology | Star, Point-to-Point. | Peer to Peer, Client Server. | Peer to peer |
| Power supply lifetime | hours, days | AA batteries; days to weeks | Solar; month to years |
| Deployment | Vehicle-placed or air-drop single sensors | Hand-emplaced | Embedded, "sprinkled" left-behind |

Table 2

ATTRIBUTES OF SENSOR NETWORK

| | |
|---|---|
| Sensors | Size: small (e.g., MEMS), large (e.g., satellites, radars)<br>Number: large, small<br>Type: passive (e.g., acoustic, video, IR), active (e.g., ladar, radar)<br>Composition or mix: heterogeneous, homogeneous<br>Spatial coverage: sparse, dense<br>Deployment: fixed, ad hoc<br>Dynamics: stationary, mobile |
| Sensing entities of interest | Extent: distributed, localized (e.g., target tracking)<br>Mobility: dynamic, static<br>Nature: cooperative (e.g., air traffic control), non-cooperative(e.g., military targets) |
| Operating location | Adverse (battlefield), Benign (factory floor). |
| Communication | Networking: wireless, wired<br>Bandwidth: Low, high |
| Processing architecture | Centralized, distributed, hybrid |
| Energy availability | Constrained (e.g., in small sensors), unconstrained (e.g., in large sensors) |

## II. SECURITY THREATS

In WSNs the security services goal is to protect resources and information from misbehavior and attack. WSN faces four basic threats, which are shown in fig. 3.

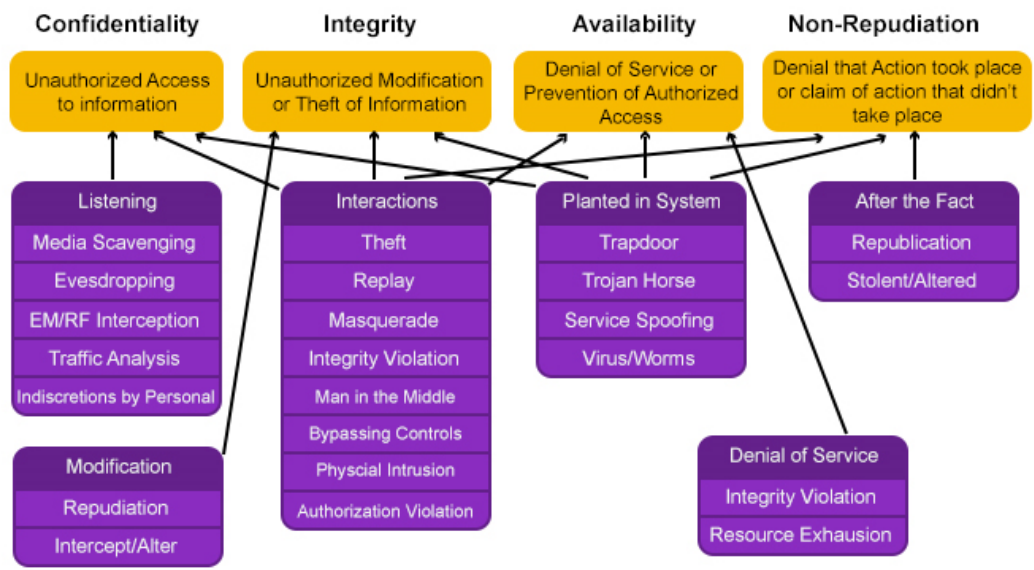

Fig. 3. Four Basic Threats

*2.1 Confidentiality*

It makes sure that only desired recipients can understand the given message. Confidentiality security countermeasures are mentioned in fig. 4.

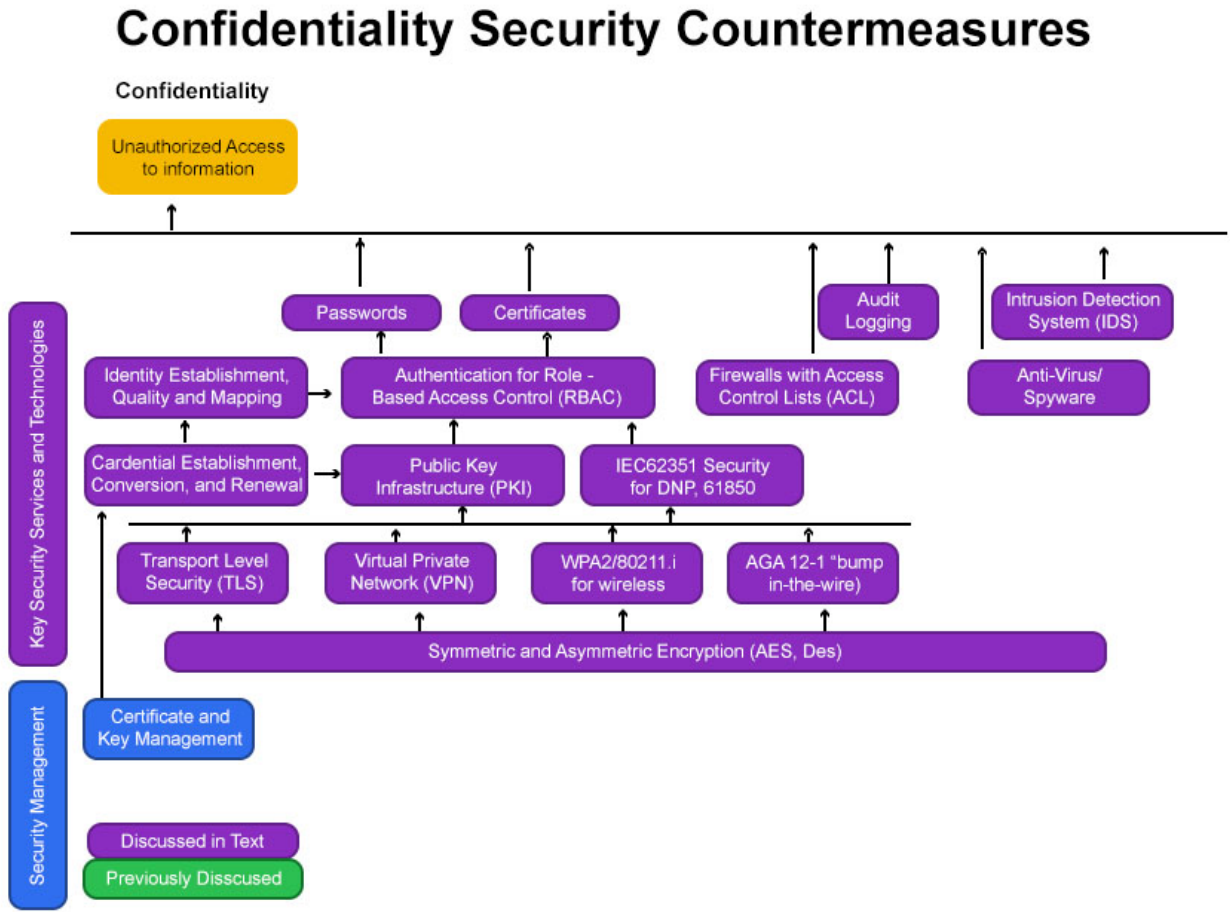

Fig. 4. Confidentiality Security and Countermeasures\

*2.2 Integrity*

It makes sure that the message is not modified by intermediate nodes which are malicious when it is sent to other nodes in a network. Countermeasures for threats on integrity are shown in fig. 5.

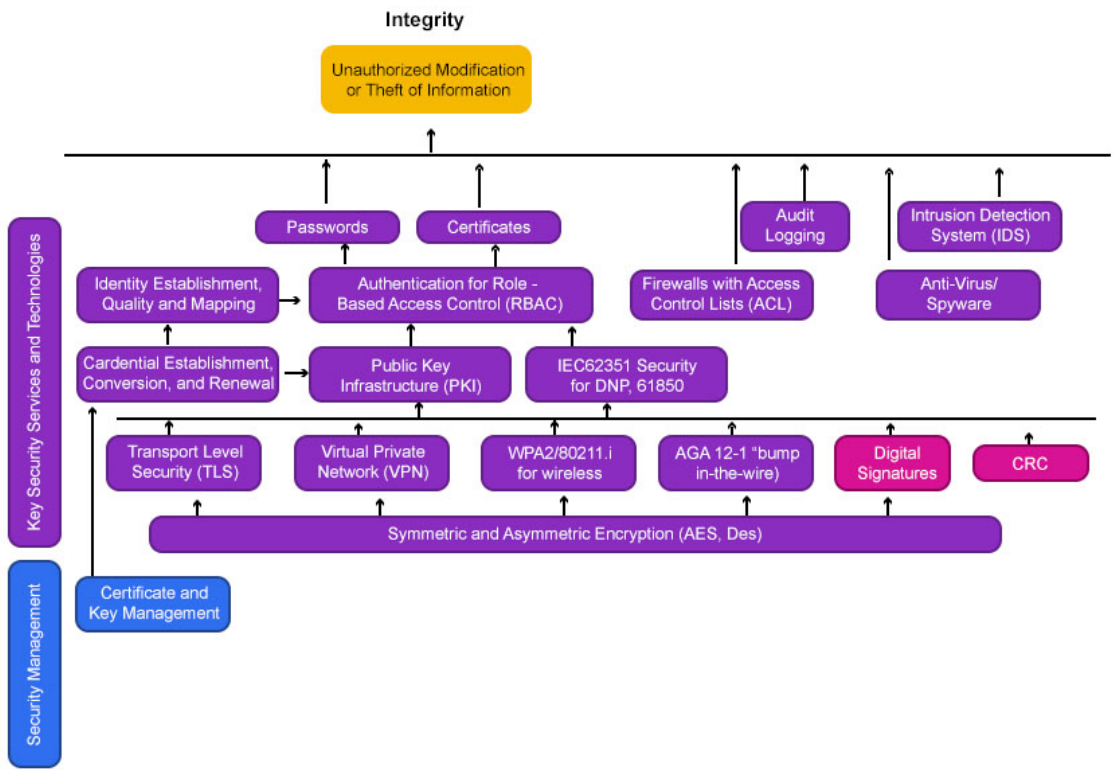

Fig. 5. Integrity Security and Countermeasures

*2.3  Availability*

It makes sure that the services of network are available which are desired even under attacks such as DOS (denial-of-service). Security countermeasures for availability threats are shown in fig. 6.

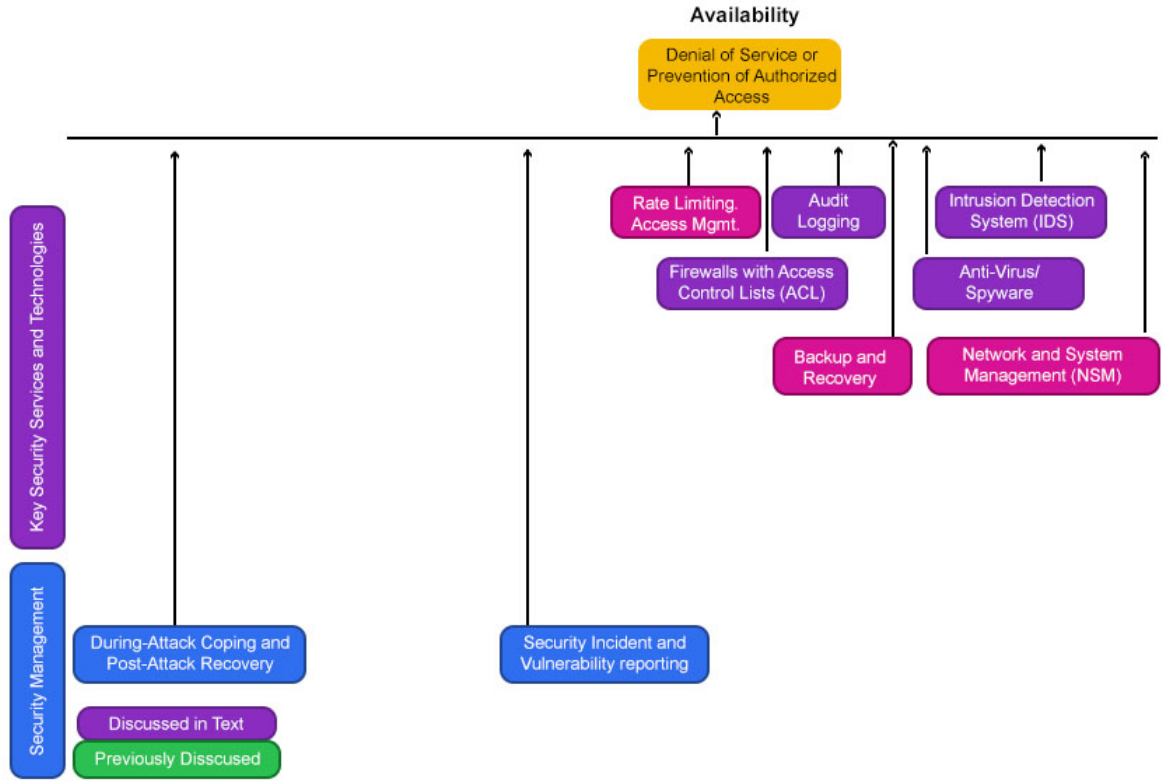

Fig. 6. Availability Security and Countermeasures

*2.4 Non-Repudiation*

Non-Repudiation refers to the facility to guarantee that a person cannot negate the authenticity of their signature. Countermeasures for Non-repudiation threats are shown in fig. 7.

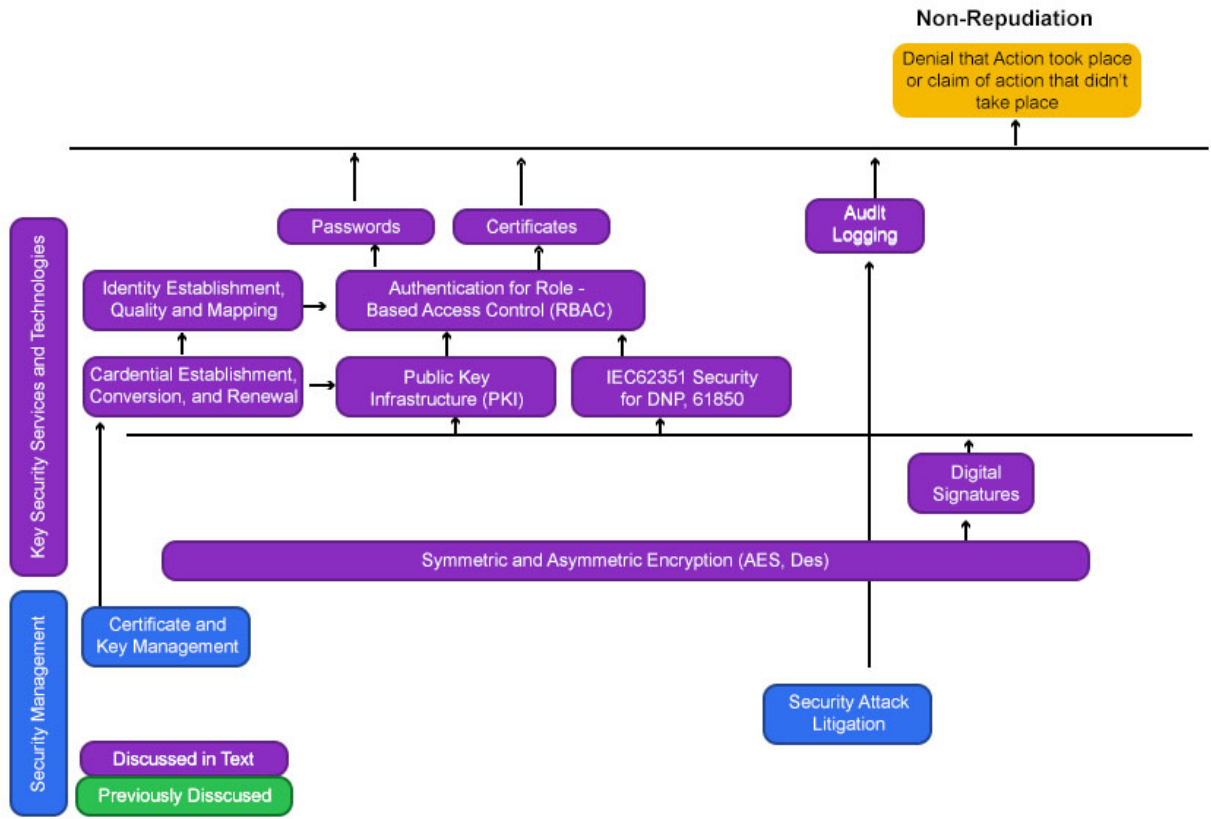

Fig. 7. Non-Repudiation Security and Countermeasures

## III. SECURITY ATTACKS

There are basically two main types of attacks that an intruder may adopt. (i) Passive attacks and (ii) Active attacks

*3.1 Passive Attacks*

These attacks are referred to just listening the communication. In this kind of attack, an intruder snoops the communications silently but does not make any changes in communication [8]. However, these attacks are normally preliminary arrangements before the active attacks. This is basically an attack against privacy.

*3.1.1 Eavesdropping*

This is very common violence against privacy. By interfering to personal data, an attacker could simply learn the content of communication. When traffic delivers control data about configuration of sensor network, which comprises hypothetically thorough information than the available through location server, snooping or eavesdropping can behave more efficiently against confidentiality.

*3.1.2 Traffic Analysis*

Messages transferred over the network are vulnerable even if they are encrypted. There is a big possibility that someone can analysis the patterns of communication. Sensor activities can possibly expose sufficient information to let an opponent to cause malevolent harm to the wireless sensor network.

*3.2 Active Attacks*

Active attacks are referred to the modifying messages and real data steam or generating the false data in communication. An intruder may repeat old data streams, changing the communicating messages or remove some selected part of important messages of communication. Some Active and passive attacks are depicted in fig. 8.

| Security Attacks | |
|---|---|
| Passive Attacks | Active Attacks |
| Traffic analysis | DoS attack |
| | Masquerade attack |
| | Replay attack |
| Eavesdropping | Selective forwarding |
| | Node replication |
| | Wormhole |
| | Sybil |
| | Sinkhole |
| | Rushing |
| | Modificatoin of messages |

Fig. 8. Active and passive attacks

*3.2.1   DoSAttack*

Denial of service (DoS) attacks are done by devastating targeted server by putting extra traffic than the server's maximum processing limit [9, 29]. DoS attack is typically performed by overwhelming the targeted node with excessive requests to overload the systems and stops all or some legitimate requests from being answered. The scheme of WSN devices typically favors reduced cost over enlarged capabilities. The basic features of sensor network devices make them susceptible to DoS attacks.

*3.2.2   Masquerade Attack*

This attack refers to the usage of a fake personality to gain the illegal access to any personal computer. An attacker acts as an unauthorized system to gain access to it or to increase greater rights than they are approved for [28]. Masquerade attack typically involves another type of active attack. For example, verification arrangements can be taken, and then an illegitimate person can obtain rights to access all the important information illegally.

*3.2.3    Replay Attack*

It usually involves passive imprisonment of the data unit and its succeeding retransmission to generate an unauthorized consequence [30, 32].This is carried out by an adversary or by an originator who interrupts the information and retransmits it.

*3.2.4    Selective Forwarding*

Timely and Secure transmission of information in the WSN is its essential requisite of the network. In this kind of attacks, malicious systems act as the normal systems and drop selected packets. In selective forwarding attack, selection of the dropping nodes can be haphazard [10].

*3.2.5    Node Replication*

In Node Replication attack, an adversary makes special easily affordable wireless sensor nodes and tricks entire network into accommodating them like the authentic nodes [12]. Node replication is difficult to detect without centralized monitoring.

*3.2.6    Wormhole Attack*

This is basically a very serious attack in which an intruder records the stream of bits or packets at specific position in wireless network and channels those to some other locations. Capturing and retransmitting of bit streams or packets could be done in selective manner [27]. Wormhole attacks normally used with the eavesdropping or selective forwarding attacks. Detection is quite problematic when used in combination with Sybil attack.

*3.2.7    Sybil Attack*

In Sybil attacks, a node presents as several duplicate nodes using the identities of other authentic nodes. Sybil attack actually goals to fault accepting schemes like distributed storage and multipath routing [11]. Sybil attack poses an important risk to "geographic routing protocols". Position knowing routing normally requires different nodes to share data with their nearby nodes to proficiently route the geographically addressed packets.

*3.2.8    Sink Hole Attack*

A malicious node represents itself as a black hole to appeal and catch all traffic especially in WSNs. An attacker snoops requests paths then shows to targeted systems that it comprises best quality or shortest distance to base station. Attacker insert itself between the collaborating nodes, it is capable to make any changes in information passing among them [26].

*3.2.9    Rushing Attack*

A latest threat that usually results in the denial-of-service (DoS) when utilized against all preceding network routing practices. An attacker distributes the malicious messages very quickly to genuine messages that reach later [22].

*3.2.10    Modification of Messages*

Message modification attacks refer to an attack where an attacker performs modifications or deletions to the content of wireless network communication. In this attack, some portion of information is transformed or real messages are postponed or recorded to produce an unauthorized effect [35].

## IV. ACTIVE ATTACKS AND THEIR COUNTERMEASURES

Countermeasures of various active attacks on wireless sensor network are mentioned in table 3.

Table 3
ACTIVE ATTACKS AND THEIR COUNTERMEASURES

| Attacks | Countermeasures |
|---|---|
| DoS Attack | Prohibit network broadcast from sensor nodes [13] |
| Masquerade Attack | RF fingerprint [34, 36] and solid authentication can be utilized to prevent Masquerade attack. |
| Replay Attack | Taxonomy of the replay attack on encryption protocols in terms of information origin and destination [33]. An effective way to evade replay attacks is by spending session tokens. |
| Selective Forwarding | Multi-path routing [13,14] and local monitoring [15] |
| Node Replication | Authentication through BS [16], Location confirmation by the witness nodes [17] and location oriented key [18] |
| Wormhole Attack | Pocket leashes [25], directional antennas [19] and topology checking by central server [20] |
| Sybil Attack | ID-based symmetric key [21] and location based key [18] |
| Sinkhole Attack | Geographic routing protocol can be one solution [14] |
| Rushing Attack | Route records by an embedding node list in ROUTERQUEST [22] |
| Modification of Messages | Link layer authentication [13, 14, 16, 23, 24] |

## V. CONCLUSION

Security is now becoming a main anxiety for Wireless Sensor Network (WSN) feature designers due to extensive safety precarious applications of these networks. In this paper, we have conferred some active attacks regarding WSN and also mentioned their best corresponding countermeasures. However, lots of open issues are still remaining to be explored. On one hand, WSN is not fully matured yet and still under progress. Numerous protocols developed until now for the WSN have not maturely satisfied the security threats. On other hand, significant protocols and requirements of WSN sometimes make it difficult to develop some solid security schemes while still supporting low operational cost. Hence, safety for the WSN is an open and fruitful research direction to be worked on.

# REFERENCES

[1] F. Akyildiz et al., "A Survey on Sensor Networks," IEEE Commun. Mag., vol. 40, no. 8, Aug. 2002, pp. 102–14.
[2] J. M. Kahn, R. H. Katz, and K. S. J. Pister, "Next Century Challenges: Mobile Networking for Smart Dust," Proc. ACM Int'l. Conf. Mobile Computing and Networking (MobiCom'99), Aug. 1999, pp. 217–78.
[3] G. J. Pottie and W. J. Kaiser, "Wireless Integrated Network Sensors," Commun. ACM, vol. 43, no. 5, May 2000, pp.51–58.
[4] Shi, Elaine, and Adrian Perrig. "Designing secure sensor networks." *IEEE Wireless Communications* 11.6 (2004): 38-43.
[5] Zhou, Yun, Yuguang Fang, and Yanchao Zhang. "Securing wireless sensor networks: a survey." *IEEE Communications Surveys & Tutorials* 10.3 (2008): 6-28.
[6] Crossbow Technology; http://www.xbow.com/, 2006.
[7] Atmel Corporation; http://www.atmel.com/, 2006.
[8] Khan, Shafiullah, et al. "Passive security threats and consequences in IEEE 802.11 wireless mesh networks." 2; 3 (2008).
[9] Sumitra, B., C. R. Pethuru, and M. Misbahuddin. "A survey of cloud authentication attacks and solution approaches." International journal of innovative research in computer and communication engineering 2.10 (2014).
[10] Sharma, Preeti, Monika Saluja, and Krishan Kumar Saluja. "A Review of Selective Forwarding attacks in Wireless Sensor Networks." *International Journal Of Advanced Smart Sensor Network Systems (IJASSN)* 2.3 (2012): 37-42.
[11] Chris Karl of, David Wagner, "Secure Routing in Wireless Sensor Networks: Attacks and Countermeasures", AdHoc Networks (elsevier) Page: 299-302, year 2003.
[12] Zhu, Wen Tao, et al. "Detecting node replication attacks in wireless sensor networks: a survey." *Journal of Network and Computer Applications* 35.3 (2012): 1022-1034.
[13] Deng, Jing, Richard Han, and Shivakant Mishra. "A performance evaluation of intrusion-tolerant routing in wireless sensor networks." *Information Processing in Sensor Networks*. Springer Berlin Heidelberg, 2003.
[14] Karlof, Chris, and David Wagner. "Secure routing in wireless sensor networks: Attacks and countermeasures." *Ad hoc networks* 1.2 (2003): 293-315.
[15] Wang, Guiling, et al. "On supporting distributed collaboration in sensor networks." *Military Communications Conference, 2003. MILCOM'03. 2003 IEEE*. Vol. 2. IEEE, 2003.
[16] Perrig, Adrian, et al. "SPINS: Security protocols for sensor networks." *Wireless networks* 8.5 (2002): 521-534.
[17] Parno, Bryan, Adrian Perrig, and Virgil Gligor. "Distributed detection of node replication attacks in sensor networks." *2005 IEEE Symposium on Security and Privacy (S&P'05)*. IEEE, 2005.
[18] Zhang, Yanchao, et al. "Location-based compromise-tolerant security mechanisms for wireless sensor networks." *IEEE Journal on Selected Areas in Communications* 24.2 (2006): 247-260.
[19] Hu, Lingxuan, and David Evans. "Using Directional Antennas to Prevent Wormhole Attacks." *NDSS*. 2004.
[20] Wang, Weichao, and Bharat Bhargava. "Visualization of wormholes in sensor networks." *Proceedings of the 3rd ACM workshop on Wireless security*. ACM, 2004.
[21] Newsome, James, et al. "The sybil attack in sensor networks: analysis & defenses." *Proceedings of the 3rd international symposium on Information processing in sensor networks*. ACM, 2004.
[22] Hu, Yih-Chun, Adrian Perrig, and David B. Johnson. "Rushing attacks and defense in wireless ad hoc network routing protocols." *Proceedings of the 2nd ACM workshop on Wireless security*. ACM, 2003.
[23] Di Pietro, Roberto, et al. "LKHW: A directed diffusion-based secure multicast scheme for wireless sensor networks." *Parallel Processing Workshops, 2003. Proceedings. 2003 International Conference on*. IEEE, 2003.
[24] Tubaishat, Malik, et al. "A secure hierarchical model for sensor network." *ACM Sigmod Record* 33.1 (2004): 7-13.
[25] Hu, Y-C., Adrian Perrig, and David B. Johnson. "Packet leashes: a defense against wormhole attacks in wireless networks." *INFOCOM 2003. Twenty-Second Annual Joint Conference of the IEEE Computer and Communications. IEEE Societies*. Vol. 3. IEEE, 2003.
[26] Ngai, Edith CH, Jiangchuan Liu, and Michael R. Lyu. "On the intruder detection for sinkhole attack in wireless sensor networks." *2006 IEEE International Conference on Communications*. Vol. 8. IEEE, 2006.
[27] Khalil, Issa, Saurabh Bagchi, and Ness B. Shroff. "LITEWORP: a lightweight countermeasure for the wormhole attack in multihop wireless networks." *2005 International Conference on Dependable Systems and Networks (DSN'05)*. IEEE, 2005.
[28] Shiu, Yi-Sheng, et al. "Physical layer security in wireless networks: a tutorial." *IEEE Wireless Communications* 18.2 (2011): 66-74.
[29] Du, Xiaojiang, and Hsiao-Hwa Chen. "Security in wireless sensor networks." *IEEE Wireless Communications* 15.4 (2008): 60-66.
[30] Syverson, Paul. "A taxonomy of replay attacks [cryptographic protocols]." *Computer Security Foundations Workshop VII, 1994. CSFW 7. Proceedings*. IEEE, 1994.
[31] Estrin, Deborah, et al. "Next century challenges: Scalable coordination in sensor networks." *Proceedings of the 5th annual ACM/IEEE international conference on Mobile computing and networking*. ACM, 1999.
[32] Das, Manik Lal. "Two-factor user authentication in wireless sensor networks." *IEEE Transactions on Wireless Communications* 8.3 (2009): 1086-1090.
[33] Syverson, Paul. "A taxonomy of replay attacks [cryptographic protocols]." *Computer Security Foundations Workshop VII, 1994. CSFW 7. Proceedings*. IEEE, 1994.
[34] Ureten, Oktay, and Nur Serinken. "Wireless security through RF fingerprinting." *Canadian Journal of Electrical and Computer Engineering* 32.1 (2007): 27-33.
[35] Beddoe, Marshall A., and Kowsik Guruswamy. "Modification of messages for analyzing the security of communication protocols and channels." U.S. Patent No. 8,601,585. 3 Dec. 2013.
[36] Padilla, J. L., et al. "RF fingerprint measurements for the identification of devices in wireless communication networks based on feature reduction and subspace transformation." Measurement 58 (2014): 468-475.